\documentclass[aps,prd,reprint,superscriptaddress,nofootinbib,10 pt]{revtex4-2}

\usepackage{amsmath,amssymb,amsfonts}
\usepackage{graphicx}
\usepackage{xcolor}
\usepackage{hyperref}
\usepackage{braket}
\newcommand{\bse}{\begin{subequations}}
\newcommand{\ese}{\end{subequations}}

\makeatletter
\renewcommand{\eqref}[1]{\textcolor{blue}{(\ref{#1})}}
\makeatother

\hypersetup{
    colorlinks=true,
    linkcolor=blue,
    citecolor=green,
    urlcolor=cyan,
    filecolor=magenta
}

\makeatletter
\def\tagform@#1{\maketag@@@{\textcolor{blue}{(\ignorespaces#1\unskip\@@italiccorr)}}}
\makeatother

\begin{document}

\title{Imaginary HTEE: Probing Criticality and Coherence}
\author{M. Ali-Akbari}
\email{m\_aliakbari@sbu.ac.ir}

\author{M. M. Daryaei Goki}
\email{m\_daryaeigoki@sbu.ac.ir}

\affiliation{Department of Physics, Shahid Beheshti University, 1983969411, Tehran, Iran}

\date{\today}

\begin{abstract}
We study the imaginary part of the holographic timelike entanglement entropy (HTEE) in a charged asymptotically AdS$_5$ black-hole background whose dual field theory exhibits a critical point in the presence of a chemical potential $\mu$ and temperature $T$. Identifying the imaginary part of the HTEE with the imaginary part of the pseudoentropy, we derive an analytical expression for this quantity and investigate its behavior on the stable and unstable thermodynamic branches. We find that the imaginary HTEE vanishes in the uncharged limit and is induced entirely by a nonzero chemical potential, signaling the emergence of quantum coherence between the initial and final states. As the dimensionless ratio $\mu/T$ approaches its critical value, we obtain an inverse-square-root divergence in the derivative of the imaginary HTEE and extract the critical exponent $\theta_I = 1/2$ analytically. We further show that the same exponent governs the temperature response at fixed chemical potential. Our results establish the imaginary part of the HTEE as a non-local quantum-information probe of criticality in strongly coupled systems, and they imply that the initial and final states become maximally coherent at the critical point.
\end{abstract}

\maketitle

\tableofcontents
\section{Introduction}

One of the most important proposals over the past two decades for describing strongly coupled field theories is the AdS/CFT correspondence, or more generally, gauge-gravity duality. It provides a powerful framework for studying such field theories by relating them to a gravitational theory in a higher dimension~\cite{Maldacena:1997re,Witten:1998qj,CasalderreySolana:2011us}. Since the gravitational theory is weakly coupled, this duality establishes a strong-weak relation between two distinct theories. By employing a suitable gravitational dual for a given physical quantity or process, its behavior can be investigated within the field theory. In general, gravitational duals admit simple interpretations, and thus the duality allows quantities that are difficult to compute directly in the strongly coupled field theory to be studied through their gravitational counterparts. The gauge-gravity duality has been extended to various setups involving finite temperature, chemical potential, and external magnetic fields, making it a versatile tool for exploring the phase structure of strongly interacting matter. In particular, holographic models have been widely used to study transport properties, such as shear viscosity and electrical conductivity, in regimes where perturbative methods break down. These developments highlight the remarkable utility of the duality as a bridge between seemingly unrelated areas of theoretical physics.

One of the particularly fruitful connections between quantum information and gravity, which has attracted a great deal of attention, is entanglement entropy~\cite{Nishioka:2009un}. In order to define it, let us consider a pure state $|\psi\rangle$ and divide it into a subregion $A$ and its complement $\bar{A}$. Hence, the density matrix of the pure state is given by $\rho = |\psi\rangle\langle\psi|$, and the reduced density matrix is $\rho_A = \mathrm{Tr}_{\bar{A}}\rho$, where the trace is taken over the complementary indices. The entanglement entropy is then defined as $S_A = -\mathrm{Tr}(\rho_A\log\rho_A)$. This quantity reflects how much information is shared between these two subsystems or, in other words, how much information is lost when the complementary system is inaccessible. Clearly, when the entanglement entropy vanishes, the two subsystems are disentangled. Entanglement entropy has emerged as a central quantity in the study of quantum many-body systems, quantum field theories, and quantum gravity. In particular, it serves as an order parameter for quantum phase transitions and characterizes the scaling of correlations in critical systems, where it typically follows a universal area law or, in one-dimensional critical systems, a logarithmic violation thereof~\cite{Nishioka:2018khk}.

In the context of gauge-gravity duality, the Ryu-Takayanagi (RT) proposal relates the holographic entanglement entropy (HEE) of a boundary region to the area of an extremal bulk surface homologous to that region~\cite{Ryu:2006bv}. Accordingly, for a spatial boundary subregion $A$, one has
\begin{equation}
S_A = \frac{\mathrm{Area}(\mathcal{A})}{4G_N},
\label{eq:RT}
\end{equation}
where $\mathcal{A}$ denotes the extremal surface in the bulk that is homologous to $A$, and $G_N$ is the Newton constant. This prescription was subsequently generalized to covariant settings by the Hubeny-Rangamani-Takayanagi proposal~\cite{Hubeny:2007xt}. This geometrical description of entanglement entropy has led to extensive studies of quantum entanglement in strongly coupled systems and has revealed that geometric quantities in the bulk can encode non-local information about the boundary quantum state~\cite{Ryu:2006ef}. Entanglement-based quantities have subsequently been employed to investigate thermalization~\cite{Balasubramanian:2011ur}, phase transitions~\cite{Klebanov:2007ws}, critical phenomena~\cite{DeWolfe:2010ca,Ebrahim:2023hru,Ebrahim:2018zcv,AliAkbari:2017pcn,Ebrahim:2019}, and other properties of strongly coupled quantum field theories~\cite{Casini:2009sr}. This remarkable connection has provided profound insights into the emergence of spacetime geometry from quantum entanglement and has motivated the celebrated conjecture that entanglement is the glue that holds spacetime together~\cite{Takayanagi:2025ula}.

Timelike entanglement entropy~\cite{Doi:2023zyl}, as a generalization of spacelike entanglement entropy, is defined as a quantity that characterizes the degree of entanglement between two systems when the subregion is temporal. It, along with its holographic dual, was introduced in~\cite{Doi:2023zyl2}, and various aspects thereof have been studied. It was shown that the HTEE is, in fact, a piecewise geodesic (spacelike plus timelike) construction, and hence the HTEE acquires a complex contribution. Most existing calculations have been performed for two-dimensional conformal field theories. To calculate the HTEE, the idea of complexified bulk geometry was then introduced in~\cite{Heller:2024}, and the same HTEE was computed within this framework. In this approach, the HTEE is given by
\begin{equation}
S_A^c = \frac{\mathrm{Area}(\mathcal{A}^c)}{4G_N},
\label{eq:HTEE}
\end{equation}
where the extremal area ${\cal A}^c$ is obtained in a complex geometry. Both of these methods have been applied to investigate various properties of the HTEE; for instance, see~\cite{Jena:2025}.

Pseudoentropy is defined via the transition matrix
\begin{equation}
\tau^{\psi|\phi} = \frac{|\psi\rangle\langle\phi|}{\langle\phi|\psi\rangle},
\label{eq:transition}
\end{equation}
which describes a transition, or a comparison, between two arbitrary pure states $|\phi\rangle$ (initial) and $|\psi\rangle$ (final) in the full Hilbert space \cite{Nakata:2020rsh,Caputa:2024zxf}. Since it is not Hermitian, $\tau^{\psi|\phi}$ is not a density matrix. The corresponding reduced transition matrix for the subregion $A$ is obtained by tracing over its complement,
\begin{equation}
\tau_A^{\psi|\phi} = \mathrm{Tr}_{\bar{A}}\bigl(\tau^{\psi|\phi}\bigr).
\label{eq:reduced_transition}
\end{equation}
The resulting reduced transition matrix is generally:
\begin{itemize}
\item \textbf{Non-Hermitian}: that is, $\tau_A^{\psi|\phi} \neq \tau_A^{\phi|\psi}$. This means the transition has a preferred direction (generally in time). The system knows whether it is going forward (from $\ket{\phi}$ to $\ket{\psi}$) or backward (from $\ket{\psi}$ to $\ket{\phi}$) between the two states. Indeed, non-Hermiticity encodes the arrow of time at the quantum level and measures the irreversibility of the transition.
\item \textbf{Not positive semi-definite}: its eigenvalues can be complex numbers.
\item \textbf{Not normalized}: although $\mathrm{Tr}(\tau_A^{\psi|\phi}) = 1$, the eigenvalues of the reduced transition matrix are not probabilities. In other words, they can be negative or greater than one.
\end{itemize}
The pseudoentropy is then defined by
\begin{equation}
S_{pA} = -\mathrm{Tr}_A(\tau_A^{\psi|\phi}\log\tau_A^{\psi|\phi}).
\label{eq:pseudoentropy}
\end{equation}
The imaginary part of the pseudoentropy is directly tied to the coherence between the initial and final states, and therefore quantifies how much quantum coherence is lost when the complement of \(A\) is traced out. It is nonzero only when there exists quantum coherence between the two states; equivalently, it serves as a direct measure of that coherence. Consequently, it attains its maximum when the states are maximally coherent. In the special
case where the two states coincide, \(|\psi\rangle = |\phi\rangle\), no coherence is present, the transition matrix reduces to an ordinary density matrix, and the pseudoentropy reduces to the standard entanglement entropy.

The imaginary contribution to the HTEE, which can be identified with pseudoentropy \cite{Doi:2023zyl2}, also exhibits a nontrivial sensitivity to the geometry of the boundary subregion. In previous studies~\cite{AliAkbari:2026,Goki:2026}, it was shown that the imaginary part of the HTEE depends on the orientation of the entangling region as well as on the anisotropy parameter of the underlying holographic background. These results indicate that (the imaginary part of) the HTEE contains information beyond the magnitude of the entanglement entropy and can be sensitive to geometric and various properties of the strongly coupled system. This motivates us to further explore whether the imaginary contribution can also encode the critical structure of the dual field theory.

In this work, we study the imaginary part of the HTEE in a charged asymptotically AdS$_5$ black hole background whose dual field theory possesses a critical point in the presence of a chemical potential $\mu$ and temperature $T$. We derive an analytical expression for the imaginary contribution to the HTEE and investigate its behavior on the stable and unstable thermodynamic branches. By studying its dependence on the dimensionless ratio $\mu/T$, we find an inverse-square-root divergence in the corresponding critical response and analytically obtain the critical exponent $\theta_I = 1/2$. Our results demonstrate that the imaginary part of the HTEE retains the leading critical structure of the underlying gravitational background and can therefore serve as a non-local probe of criticality.

\section{Holographic Setup and Thermodynamic Phase Structure}

We are interested in exploring the behavior of the HTEE near a critical point in a strongly coupled thermal field theory using the gauge-gravity duality. To this end, we consider a five-dimensional charged asymptotically AdS black hole, which is dual to a four-dimensional field theory exhibiting a critical point. This geometry provides a convenient holographic realization of a field theory with a nontrivial phase structure and will serve as the gravitational background throughout our analysis. The five-dimensional background metric is given by~\cite{Sahay:2010rq}
\begin{equation}
ds^2 = e^{2A(r)}\left(-h(r)dt^2 + d\vec{x}^2\right) + \frac{e^{2B(r)}}{h(r)}dr^2,
\label{eq:metric}
\end{equation}
where the three spatial directions are denoted by $\vec{x} = (x,y,z)$ and the metric components are
\begin{equation}
\begin{aligned}
A(r) &= \ln r + \frac{1}{6}\ln\left(1 + \frac{Q^2}{r^2}\right), \\
B(r) &= -\ln r - \frac{1}{3}\ln\left(1 + \frac{Q^2}{r^2}\right), \\
h(r) &= 1 - \frac{M^2}{r^2(r^2 + Q^2)}, \\
\phi(r) &= -\frac{2}{3}\ln\left(1 + \frac{Q^2}{r^2}\right), \\
A_t(r) &= \left(\frac{-MQ}{r^2 + Q^2} + \frac{MQ}{r_h^2 + Q^2}\right),
\end{aligned}
\label{eq:metric_components}
\end{equation}
where $r$ is the radial coordinate and the boundary is located at $r \to \infty$. Throughout this work, we set the AdS radius to unity. The event horizon is determined by the blackening function vanishing, $h(r_h) = 0$, which yields:
\begin{equation}
r_h = \sqrt{\frac{\sqrt{Q^4 + 4M^2} - Q^2}{2}}.
\label{eq:horizon}
\end{equation}
The temporal component of the gauge field, $A_t$, is chosen such that it vanishes at the event horizon, ensuring regularity, while approaching a finite value at the asymptotic boundary. The field $\phi$ is the dilaton, and the parameters $M$ and $Q$ characterize the mass and charge of the black hole, respectively.

The thermodynamic quantities of the dual field theory can be obtained from the standard holographic dictionary. In particular, the chemical potential and temperature in the strongly coupled field theory are given by~\cite{Sahay:2010rq}:
\begin{equation}
\begin{aligned}
\mu &= \frac{Qr_h}{\sqrt{Q^2 + r_h^2}}, \\
T &= \frac{Q^2 + 2r_h^2}{2\pi\sqrt{Q^2 + r_h^2}}.
\end{aligned}
\label{eq:thermo}
\end{equation}
The thermodynamic structure of the background becomes particularly transparent when expressed in terms of the dimensionless ratio $Q/r_h$. Using the relations between $\mu$, $T$, $Q$, and $r_h$, this ratio can be written in terms of the dimensionless boundary quantity $\mu/T$ as:
\begin{equation}
\frac{Q}{r_h} = \sqrt{2}\left(\frac{1\pm\sqrt{1 - \left(\frac{\sqrt{2}}{\pi}\frac{\mu}{T}\right)^2}}{\frac{\sqrt{2}}{\pi}\frac{\mu}{T}}\right).
\label{eq:Q_rh}
\end{equation}
It was shown in~\cite{DeWolfe:2010ca2} that there is a critical point located at $(Q/r_h)_c = \sqrt{2}$ or equivalently $(\mu/T)_c = \pi/\sqrt{2}$. In other words, for a given value of $\mu/T$ below its critical value, this relation admits two distinct values of $Q/r_h$, corresponding to the $\pm$ signs in~\eqref{eq:Q_rh}. These two solutions correspond to two thermodynamic branches of the black hole background. The branch satisfying $Q/r_h < \sqrt{2}$, corresponding to the minus sign in~\eqref{eq:Q_rh}, is thermodynamically stable, whereas the other branch corresponds to the thermodynamically unstable solution. The coexistence of these two branches is characteristic of a first-order phase transition in the dual field theory. The two branches continuously approach each other and merge at the critical point. Thus, the stable and unstable branches become indistinguishable at the critical point. This thermodynamic structure motivates studying the HTEE in the vicinity of the critical point, aiming to examine whether the entanglement structure reflects the underlying phase behavior.

\section{Holographic Timelike Entanglement Entropy}
To evaluate the HTEE, we consider a timelike interval on the boundary extending along the time direction while keeping the spatial coordinate \(x = 0\) fixed, as
\begin{equation}
A:\left\{-\frac{\Delta t}{2} < t < \frac{\Delta t}{2}, -\frac{L}{2} < y,z < \frac{L}{2}, x = 0\right\},
\label{eq:boundary_region}
\end{equation}
where $L \to \infty$ denotes the transverse directions. The corresponding extremal surface in the bulk geometry is parametrized by $t(r)$ and the area functional is therefore:
\begin{equation}\begin{split}
\mathcal{A} &= V_{yz}\int dr\, e^{2A(r)}\sqrt{\frac{e^{2B(r)}}{h(r)} - e^{2A(r)}h(r)t'(r)^2}, \cr
&\equiv V_{yz}\int dr\,\mathcal{L},
\label{eq:area_functional}
\end{split}\end{equation}
where $V_{yz} = \int dy\,dz$. It is clear that the coordinate $t(r)$ does not appear explicitly in the Lagrangian, and therefore $t(r)$ is cyclic. Consequently, the associated momentum is conserved:
\begin{equation}
E = \frac{\partial\mathcal{L}}{\partial t'},
\label{eq:momentum}
\end{equation}
where $E$ is the conserved quantity associated with translations along the time direction. The profile $t(r)$ is then obtained by solving the above conservation equation. After some straightforward algebra, the conservation equation~\eqref{eq:momentum} can be solved for the profile of the extremal surface, yielding:
\begin{equation}
t'(r) = \frac{Ee^{(B - A)}}{h\sqrt{e^{6A}h + E^2}}.
\label{eq:t_prime}
\end{equation}
Normally, the turning point of the extremal surface, located at $x = 0$ by symmetry, can be obtained from $t'(r = r_*) \to \infty$, or equivalently:
\begin{equation}
e^{6A(r_*)}h(r_*) + E^2 = 0,
\label{eq:turning_point_condition}
\end{equation}
leading to smooth behavior at this point. However, due to the timelike nature of the surface, this equation clearly shows that there is no real critical point satisfying~\eqref{eq:turning_point_condition}, forcing us to use complex numbers. Substituting~\eqref{eq:t_prime} into~\eqref{eq:area_functional}, it is straightforward to obtain the area functional. However, the explicit form of this integral is non-trivial and will be addressed below.

\begin{figure}[htbp]
\centering
 \includegraphics[width=0.48\textwidth]{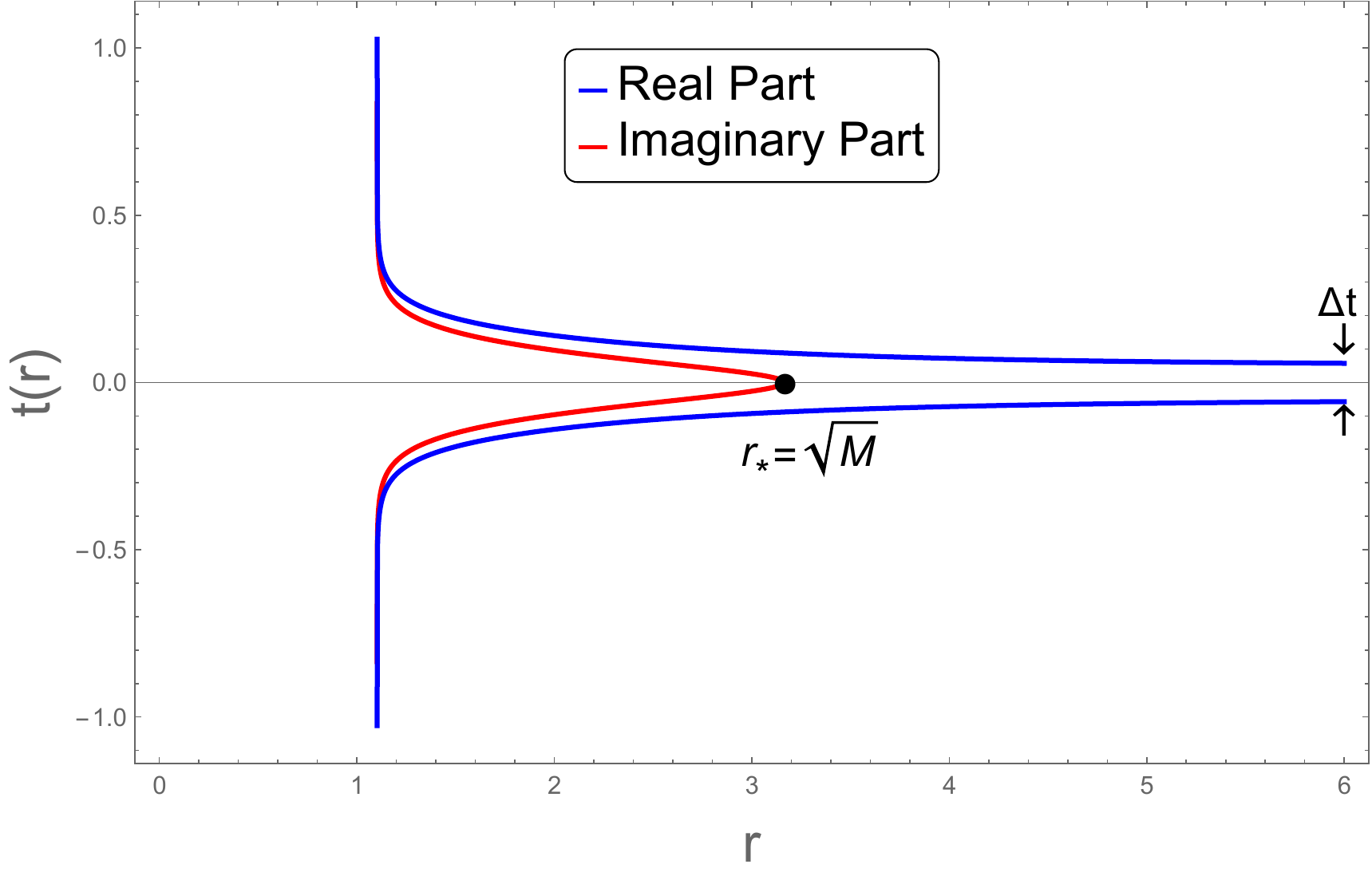}
 \caption{The real (blue) and imaginary (red) parts of the 
extremal-surface profile $t(r)$ as functions of the radial 
coordinate $r$, for $\Delta t = 0.1$, $Q = 9$, $M = 10$, and 
$E = 90(={\cal E}=MQ)$. The imaginary part diverges at the turning point $r_*$, where the square root in \eqref{19-2} vanishes, while 
the real part remains smooth. The imaginary part of $t(r)$ is 
nonzero only in the region $r_h < r < r_*$.}
\label{1}
\end{figure}

Hereafter, we have two options to compute the HTEE:
\begin{enumerate}
\item Using the idea of a complexified bulk geometry introduced in~\cite{Heller:2024}, by evaluating the integral,
\begin{equation}
\frac{\mathcal{A}^c}{2} = V_{yz}\int_{r_*}^{\infty}dr\,\frac{e^{5A(r) + B(r)}}{\sqrt{e^{6A(r)}h(r) + E^2}}.
\label{eq:complex_area}
\end{equation}
By solving this integral analytically, one can obtain the real and imaginary parts of the complex extremal surface, which correspond to the real and imaginary parts of the HTEE, respectively. Unfortunately, this integral cannot be evaluated analytically. Moreover, note that the time interval on the boundary is obtained from
\begin{equation}
\frac{\Delta t}{2} = \int_{r_*}^{\infty}t'(r)\,dr.
\label{eq:boundary_time}
\end{equation}
Note that the constant \(E\) is related to the turning point through \eqref{eq:turning_point_condition}.

\item The other method, introduced and used in~\cite{Afrasiar:2025b}, involves working with the real bulk geometry and decomposing the extremal surface into two parts corresponding to $E = i\mathcal{E}$ and $E = \mathcal{E}$ for which 
\begin{equation}
E=i\sqrt{e^{6A(r_*)}h(r_*)}\equiv i{\cal E}.
\end{equation}
The extremal surface area is then a sum of the following two contributions:
\bse\label{eq:decomposed_area}
\begin{align}
\label{19-1} \frac{\mathcal{A}_{\mathrm{R}}}{2} &\equiv \frac{\mathrm{Re}\mathcal{A}}{2} = V_{yz}\int_{r_h}^{\infty}dr\,\frac{e^{5A + B}}{\sqrt{e^{6A}h + \mathcal{E}^2}},  \\
\label{19-2}\frac{i\mathcal{A}_{\mathrm{I}}}{2} &\equiv \frac{i\mathrm{Im}\mathcal{A}}{2} = V_{yz}\int_{r_*}^{r_h}dr\,\frac{e^{5A + B}}{\sqrt{e^{6A}h - \mathcal{E}^2}},
\end{align}\ese
with $E={\cal E}$ and $E=i{\cal E}$, respectively. As a result, we have $\mathcal{A} = \mathcal{A}_R + i\mathcal{A}_I$. In this method, the integrals for the real and complex parts are different, which allows us to evaluate the integral \eqref{19-2} analytically, as we will see, although the real part remains difficult to solve analytically. Since the entanglement entropy is related to the real integral~\eqref{19-1} has been studied in detail in the literature, we focus on the complex part. The time interval on the boundary is given by:
\begin{equation}
\frac{\Delta t}{2} = \int_{r_*}^{r_h}t_I'(r)\,dr + \int_{r_h}^{\infty}t_R'(r)\,dr.
\label{eq:boundary_time_2}
\end{equation}
For more illustration, we plot the configuration $t(r)$ in terms of radial coordinate in figure \ref{1}.
\end{enumerate}

The point which is worthwhile to point out here is that in
\eqref{eq:turning_point_condition}, for the first method, one can obtain the turning point \(r_*\) in terms of \(E\), and it generally turns out that the turning point is purely imaginary, that is \(r_* = \pm i|r_*|\). The only difference between the plus and minus sign appears in the imaginary parts of the HTEE, whose signs flip from negative (positive) to positive (negative).
However, for the second approach, the turning point is real, as one can see from \eqref{eq:turning_point}, and this ambiguity does not appear.

After calculating the extremal surface area $\mathcal{A}$, either analytically or numerically, one substitutes it into the RT formula~\eqref{eq:RT}, and the HTEE is obtained.

\subsection{Analytical Evaluation of the Imaginary Part}

For a generic value of $\mathcal{E}$, the integrals in~\eqref{19-2} cannot be evaluated in closed form. However, a particular choice of $\mathcal{E}$ may render the integral analytically tractable. Motivated by this, we select
\begin{equation}
\mathcal{E} = MQ.
\label{eq:choice_E}
\end{equation}
Using $E=i{\cal E}$ and substituting this value into the turning-point condition~\eqref{eq:turning_point_condition}, we obtain
\begin{equation}\begin{split}
0 &= e^{6A(r_*)}h(r_*) - M^2Q^2 \cr
&= (r_*^2 + Q^2)(r_*^2 - M)(r_*^2 + M),
\label{eq:turning_point_factorized}
\end{split}\end{equation}
where the factorization follows from the explicit forms of $A(r)$ and $h(r)$. Since the turning point must be real and positive, and lies in the range $r_h < r_* < \infty$, the only admissible solution is
\begin{equation}
r_* = \sqrt{M}.
\label{eq:turning_point}
\end{equation}
It now remains to verify that this turning point indeed lies outside the horizon. From the horizon condition $h(r_h) = 0$, we have $r_h^2(r_h^2 + Q^2) = M^2$ so that
\begin{equation}
r_*^2 = M = r_h\sqrt{r_h^2 + Q^2}.
\label{eq:r_star_relation}
\end{equation}
For any $Q > 0$, it follows immediately that $r_* > r_h$, which is consistent with the requirement that the extremal surface extends from the turning point to the horizon. In the special case $Q = 0$, the same expression gives $r_* = r_h$ which corresponds to a degenerate limit. Thus, for any non-zero charge, the chosen value $\mathcal{E} = MQ$ yields a valid and well-defined turning point.

After substituting the metric components~\eqref{eq:metric_components}, the imaginary part of the area functional takes the form
\begin{equation}
\frac{i{\cal A}_I}{2} = V_{yz}\int_{r_*}^{r_h}\frac{r^3\sqrt{r^2 + Q^2}}{\sqrt{E^2 - M^2r^2 + Q^2r^4 + r^6}}\,dr.
\label{eq:A_I_integral}
\end{equation}
Using $\mathcal{E} = MQ$ and the factorization in~\eqref{eq:turning_point_factorized}, all explicit $Q$-dependent factors cancel, and the area functional reduces to
\begin{equation}
\frac{{\cal A}_I}{2} = -V_{yz}\int_{\sqrt{M}}^{r_h}\frac{r^3\,dr}{\sqrt{M^2 - r^4}},
\label{eq:A_I_reduced}
\end{equation}
but note that the implicit $Q$ dependence exists in $r_h$. The remaining integral is elementary and evaluates to
\begin{equation}
{\cal A}_I = V_{yz}\sqrt{M^2 - r_h^4}.
\label{eq:A_I_result}
\end{equation}
Using \eqref{eq:RT}, we define the normalized imaginary HTEE density as
\begin{equation}
\begin{aligned}
\mathrm{Im}\left[s_{\Delta t}\right] &\equiv \mathrm{Im}\left[\frac{4G_N}{V_{yz}}S_{\Delta t}\right] = \sqrt{M^2 - r_h^4}\\
& = Qr_h 
= \frac{\pi\mu T}{2}\left(3 \pm \sqrt{1 - \frac{2\mu^2}{\pi^2T^2}}\right).
\end{aligned}
\label{eq:Im_HTEE}
\end{equation}
This analytical expression gives the imaginary part of the HTEE directly in terms of the horizon radius and \(r_*\), with the latter related to the boundary condition \(\Delta t\) through~\eqref{eq:boundary_time_2}. However, because the real part of the extremal surface cannot be obtained analytically, even for the case of \(E=\mathcal{E}\), a closed-form expression for this dependence on \(\Delta t\) is unavailable. We will use
the final result~\eqref{eq:Im_HTEE} in the next section to study the behavior of the imaginary HTEE near the critical point.

For this specific case, a few points are worth emphasizing.

\begin{itemize}
\item It is well known that the imaginary part of the HTEE vanishes for the five-dimensional thermal conformal field theory (i.e., when $Q = 0$). Our result, however, reveals that a non-zero chemical potential gives rise to a non-vanishing imaginary contribution. In the language of pseudoentropy, the introduction of these new degrees of freedom makes the reduced transition matrix non-Hermitian, thereby allowing the HTEE to serve as a genuine probe of pseudoentropy.

\item Our final result~\eqref{eq:Im_HTEE} has mass dimension 2. In the conformal limit, a dimension-2 quantity at finite temperature and chemical potential must scale as $[{\rm energy}]^2$ times a function of the dimensionless ratio $\mu/T$. Equation~\eqref{eq:Im_HTEE} exhibits exactly this structure, and we may therefore write
\begin{equation}
\frac{\mathrm{Im}\left[s_{\Delta t}\right]}{T^2} = \frac{\pi}{2}\frac{\mu}{T}\left(3 \pm \sqrt{1 - \frac{2\mu^2}{\pi^2T^2}}\right),
\label{eq:Im_HTEE_conformal}
\end{equation}
which is consistent with the conformal symmetry of the underlying theory.

\item It is also instructive to examine the small-$\mu/T$ limit of the imaginary HTEE. Expanding~\eqref{eq:Im_HTEE} for $\mu/T \ll 1$, we obtain
\begin{equation}\begin{split}
&\mathrm{Im}\left[s_{\Delta t}\right] =\cr \quad\quad & \quad\quad \begin{cases}
\pi\mu T + \frac{\mu^3}{2\pi T} + \mathcal{O}\left(\frac{\mu^5}{T^3}\right), & \text{(stable)}, \\
2\pi\mu T - \frac{\mu^3}{2\pi T} + \mathcal{O}\left(\frac{\mu^5}{T^3}\right), & \text{(unstable)}.
\end{cases}
\label{eq:small_mu_expansion}
\end{split}\end{equation}
Thus, in the small-$\mu/T$ regime the imaginary HTEE vanishes, in contrast to the real part, which approaches its uncharged value. The two thermodynamic branches approach the conformal limit with different slopes, $\pi$ and $2\pi$ for the stable and unstable branches, respectively, reflecting their distinct responses to a small chemical potential. Equivalently, in terms of the dimensionless ratio,
\begin{equation}
\frac{\mathrm{Im}\left[s_{\Delta t}\right]}{T^2} \simeq \begin{cases}
\pi\frac{\mu}{T}, & \text{(stable)}, \\
2\pi\frac{\mu}{T}, & \text{(unstable)},
\end{cases}
\label{eq:small_mu_ratio}
\end{equation}
so that the imaginary HTEE is entirely induced by the chemical potential and vanishes when $\mu = 0$, consistent with the uncharged theory. The expansion is regular in $\mu/T$, with no singularity at $\mu/T = 0$; the critical point at $(\mu/T)_c = \pi/\sqrt{2}$ lies far from this regime and does not affect the leading behavior.

\item The imaginary part of the HTEE is strictly positive throughout the physical domain. For the stable branch, the term $3 - \sqrt{1 - 2\mu^2/(\pi^2T^2)}$ ranges from 2 to 3 as the critical point is approached from below, and the overall prefactor $\pi\mu T/2$ is positive for $\mu, T > 0$. The unstable branch yields an even larger positive value, so the positivity holds for both branches.

\item For the thermodynamically stable branch, the function $3 - \sqrt{1 - 2\mu^2/(\pi^2T^2)}$ increases monotonically with $\mu/T$ and attains its maximum exactly at the critical point $(\mu/T)_c = \pi/\sqrt{2}$. Consequently, the imaginary HTEE reaches its maximal value $\mathrm{Im}[s_{\Delta t}]_{\mathrm{max}} = \frac{3\pi}{2}\mu_cT_c$ precisely at the critical point. Thus, at least within our specific model and for the chosen stable branch, the HTEE identifies the critical point as the point of maximal imaginary contribution.

\item Finally, we note that the imaginary part of the HTEE is independent of the time interval \(\Delta t\), as can be seen from \eqref{eq:Im_HTEE}. This is not a new result: for instance, in two-dimensional conformal field theory the imaginary part equals \(i\pi c/6\), where \(c\) is the central charge. Moreover, in four-dimensional conformal field theory it vanishes, while in the presence of an anisotropy parameter \(a\) it is proportional
to \(a^2\) and again independent of the time interval. For this reason, our discussion does not depend on the choice of the time interval.
\end{itemize}

According to our setup, one may redefine the black hole parameters as
\begin{equation}
(M, Q, r_h) \longrightarrow (\lambda^2M, \lambda Q, \lambda r_h),
\label{eq:rescaling}
\end{equation}
with $\lambda > 0$. Under this rescaling, the relevant quantities transform as
\begin{equation}\begin{split}
&\mu \to \lambda\mu, \quad T \to \lambda T, \quad r_* \to \lambda r_*,\cr 
&\mathcal{E} \to \lambda^3\mathcal{E}, \quad \mathcal{A}_I \to \lambda^2\mathcal{A}_I.
\label{eq:rescaling_quantities}
\end{split}\end{equation}
This redefinition preserves the location of the critical point, leaves the dimensionless ratio $\mu/T$ invariant, and, as we will show, does not alter the critical exponent. Consequently, it generates a one-parameter family of solutions labelled by $\lambda$, whose imaginary part scales as $\lambda^2$. 
The rescaling \eqref{eq:rescaling} changes the numerical value of the imaginary part of the HTEE by a factor of \(\lambda^2\), reflecting the fact that \(\mathrm{Im}[s_{\Delta t}]\) has mass dimension 2. However, the dimensionless coherence \(\mathrm{Im}[s_{\Delta t}]/T^2\) is invariant under this rescaling
and depends only on \(\mu/T\). Since the two systems related by the rescaling have the same value of \(\mu/T\), they are physically equivalent, the rescaling is a conformal transformation, not a physical change.

\subsection{Numerical Verification}
\begin{figure}[htbp]
\centering
\includegraphics[width=0.48\textwidth]{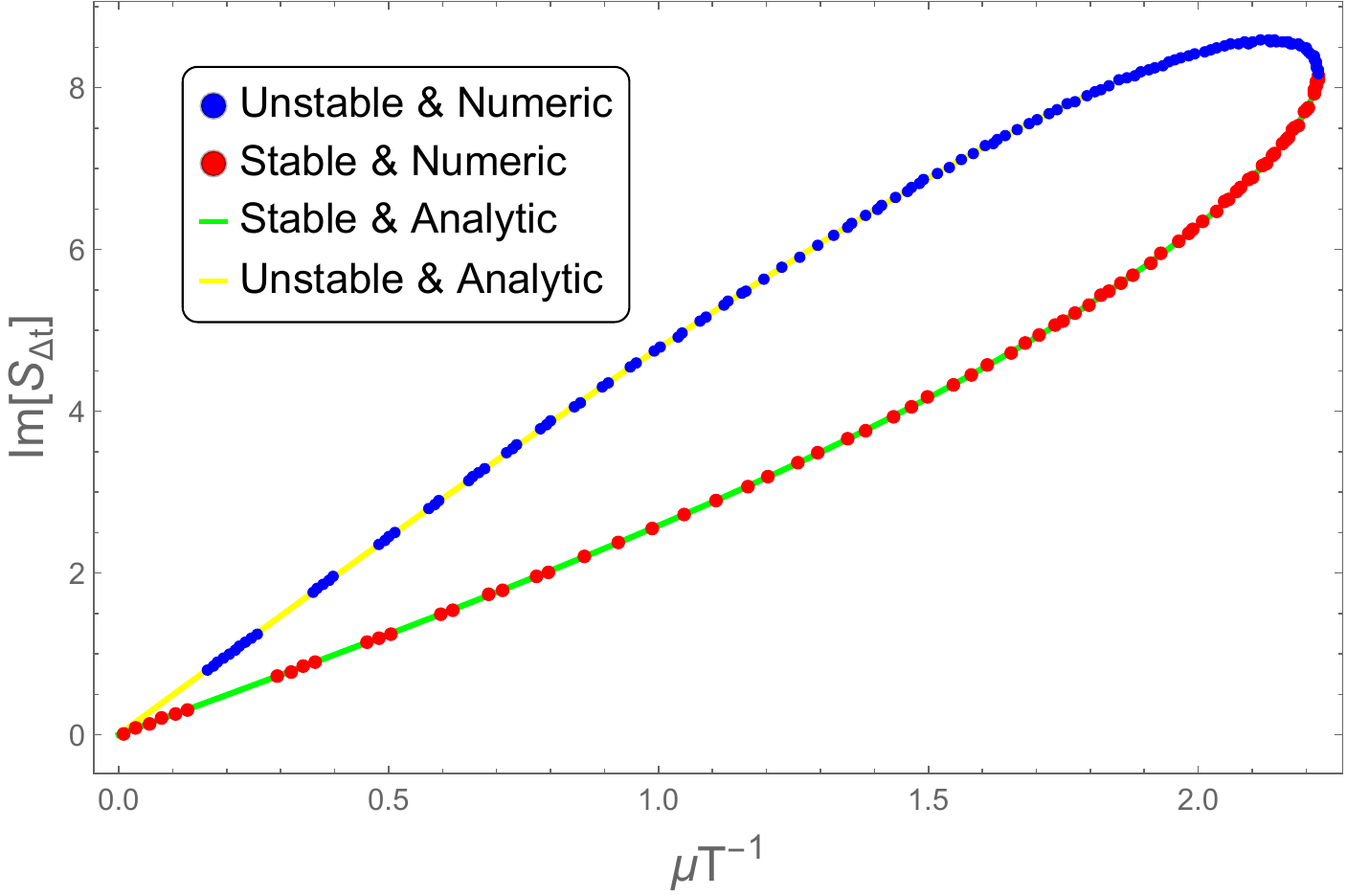}
\caption{The imaginary part of the HTEE \eqref{19-2} as a function of $\mu/T$, with $T=1.25$ and $\Delta t=0.03$ held fixed. Note that each point in this figure corresponds to a solution with a different value of $\mathcal{E}=MQ$. Alternatively, one may keep the constant $\mathcal{E}$ and the interval $\Delta t$ fixed, in which case the same behavior is observed.}
\label{2}
\end{figure}

\begin{figure}[htbp]
\centering
 \includegraphics[width=0.48\textwidth]{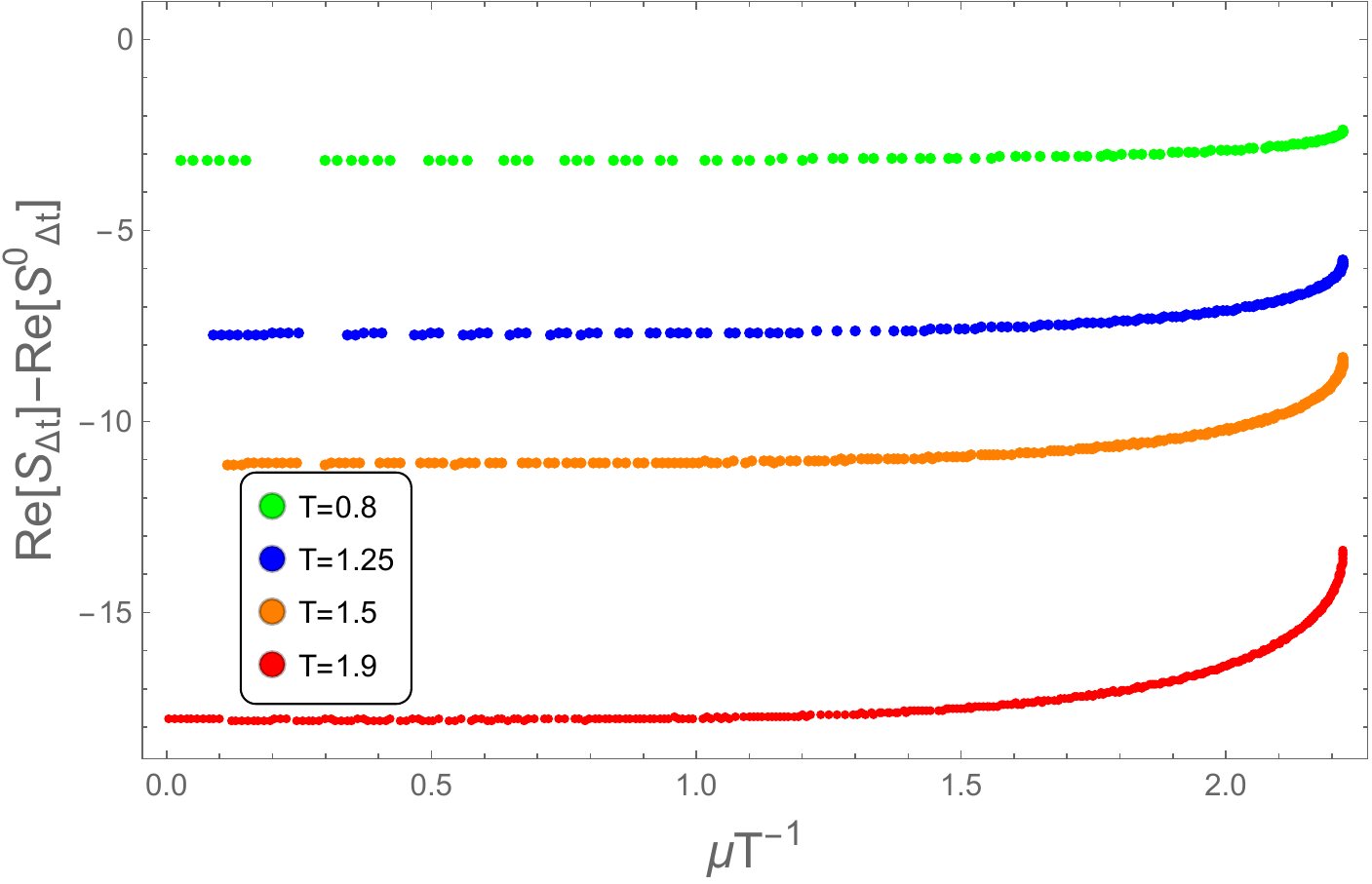}
 \caption{The real part of the HTEE is plotted as a function of $\mu T^{-1}$ for four different temperatures using \eqref{19-1} along the stable branch. $S^0_{\Delta t}$ denotes the real part of the HTEE in the vacuum AdS background, corresponding to $M=0$ and $Q=0$.}
\label{3}
\end{figure}

As mentioned earlier, obtaining the real and imaginary parts of the HTEE analytically is generally impossible. In the previous subsection, for a specific value of the constant $\mathcal{E}$, the analytic result was obtained and discussed only for the imaginary part. To crosscheck our analytical result and also investigate the solution for various values of the constant $\mathcal{E}$, in the following we compute the integrals \eqref{eq:decomposed_area} numerically and report the results.

In figure \ref{2}, the imaginary part of the HTEE, for the stable and unstable branches, has been plotted in terms of the dimensionless parameter $\mu/T$. As this figure shows, the numerical solution for $\mathcal{E}=MQ$ is exactly consistent with our analytical result. In fact, the dotted red and blue points in this figure, which are the solutions obtained from \eqref{19-2}, lie exactly on the curves plotted from \eqref{eq:Im_HTEE} for both the stable and unstable branches. The figure indicates that the imaginary part starts from zero at $\mu=0$ and then increases monotonically up to its maximum value at the critical point. Moreover, the imaginary part of the HTEE is naturally interpreted in the framework of pseudoentropy and corresponds to the imaginary part of the pseudoentropy. With this identification, our result implies that the initial and final states are maximally coherent at the critical point. This establishes the imaginary of the HTEE as a genuine probe of the non-Hermitian structure of the reduced transition matrix in a strongly coupled system with a critical point.

According to our setup, the initial and final states are identified with the AdS$_5$ black hole and the charged AdS$_5$ black hole, respectively. We then show that the imaginary part of the HTEE, or equivalently the pseudoentropy, vanishes when the charge $Q$, or equivalently $\mu$, is zero. This means that there is no coherence between two AdS$_5$ black holes, or equivalently between two thermal states, independent of the temperature. Upon turning on the chemical potential, coherence appears and, at least in our model, a nonzero chemical potential induces coherence. As a result, the temperature and the ratio $\mu/T$ control how much quantum coherence is lost.

For the stable branch, in figure \ref{3}, the real part of the HTEE, obtained from \eqref{19-1}, is plotted in terms of \(\mu/T\). This behavior has already been reported in \cite{Amrahi:2020jqz}, and its properties have been discussed.

\section{Critical Behavior of the Imaginary HTEE}
In order to study the behavior of the imaginary part of the HTEE near the critical point, it is useful to write~\eqref{eq:Im_HTEE} as
\begin{equation}
\frac{\mathrm{Im}\left[s_{\Delta t}\right]}{T^2} = \frac{\pi}{2}x\left(3 \pm \sqrt{1 - \left(\frac{x}{x_c}\right)^2}\right),
\label{eq:Im_HTEE_x}
\end{equation}
where $x \equiv \frac{\mu}{T}$ and $x_c \equiv (\frac{\mu}{T})_c$. The critical behavior can therefore be extracted directly from the analytical structure of~\eqref{eq:Im_HTEE_x}. At fixed temperature, we thus have
\begin{equation}\begin{split}
\frac{\partial}{\partial x}\left(\frac{\mathrm{Im}\left[s_{\Delta t}\right]}{T^2}\right)_T
= \frac{\pi}{2}\Bigg[3 & \pm \sqrt{1-\left(\frac{x}{x_c}\right)^2}\cr
&\mp \frac{x^2/x_c^2}{\sqrt{1-\left(\frac{x}{x_c}\right)^2}}\Bigg].
\label{eq:derivative}
\end{split}\end{equation}
In the vicinity of the critical point, $x \to x_c$, the second term dominates and yields
\begin{equation}
\left(\frac{\partial\mathrm{Im}\left[s_{\Delta t}\right]}{\partial x}\right)_T \sim \mp \frac{\pi}{2}\,T^2 x_c\left(x_c^2-x^2\right)^{-1/2}.
\label{eq:derivative_critical}
\end{equation}
Hence,
\begin{equation}
\left(\frac{\partial {\rm Im}\left[s_{\Delta t}\right]}{\partial(\mu/T)}\right)_T \sim \left[\left(\frac{\mu}{T}\right)_c - \frac{\mu}{T}\right]^{-\theta_I}, \quad \theta_I = \frac{1}{2}.
\label{eq:critical_exponent}
\end{equation}
Both branches exhibit the same critical exponent. Thus, the exponent $\theta_I = 1/2$ follows directly from the square-root branch structure of the imaginary HTEE. This result is consistent with previous holographic studies showing that entanglement-based observables can detect critical points and inherit the characteristic scaling of the underlying gravitational background. In particular, HEE and the entanglement of purification have been shown to exhibit the same behavior near the critical point \cite{Amrahi:2020jqz}. Our result extends this picture to the imaginary contribution of HTEE. This establishes the imaginary HTEE as a non-local quantum-information probe of the critical behavior of the dual field theory.

We can further characterize this critical behavior by considering the temperature response at fixed chemical potential. In analogy with the thermodynamic heat capacity, we define
\begin{equation}
C_I \equiv T\left(\frac{\partial \mathrm{Im}\left[s_{\Delta t}\right]}{\partial T}\right)_\mu.
\label{eq:C_I}
\end{equation}
Here, $C_I$ should not be interpreted as a thermodynamic heat capacity. Rather, it is an entanglement response function that characterizes the sensitivity of the imaginary part of the HTEE to temperature variations at fixed chemical potential. The singular behavior of $C_I$ is directly inherited from that of $\partial{\rm Im} [s_{\Delta t}]/\partial x$. We then obtain
\begin{equation}
C_I \sim \pm \frac{\pi^2}{2\sqrt{2}}\,T_c^3\left(T^2 - T_c^2\right)^{-1/2}.
\end{equation}
Consequently,
\begin{equation}
C_I \sim |T - T_c|^{-\alpha_I}, \qquad \alpha_I = \frac{1}{2}.
\label{eq:alpha_I}
\end{equation}
Thus, the imaginary HTEE exhibits the same inverse-square-root singularity when the critical point is approached through the temperature direction. The agreement between the $\mu/T$ response and the temperature response demonstrates that the critical behavior is not an artifact of a particular parametrization, but rather reflects the underlying square-root structure of the thermodynamic branches. In this sense, the imaginary HTEE provides a non-local quantum-information observable that retains the leading critical scaling of the holographic background.

An important question is whether the critical exponent $\theta_I = 1/2$ is probe-independent. In our previous studies~\cite{AliAkbari:2026}, we found that the imaginary HTEE depends on the orientation of the entangling region and on the anisotropy parameter of the background. However, the critical exponent obtained here follows directly from the square-root branch structure of the thermodynamic quantities, which is a property of the background geometry rather than of the particular probe. We therefore expect $\theta_I = 1/2$ to be universal across different probes.
%
%
%


\section*{Acknowledgment}
This work is based upon research funded by Iran National Science Foundation (INSF) under project No 40509128.


\begin{thebibliography}{99}

\bibitem{Maldacena:1997re}
J.~M.~Maldacena,
``The Large N limit of superconformal field theories and supergravity,''
Adv.\ Theor.\ Math.\ Phys.\ \textbf{2}, 231-252 (1998)
[arXiv:hep-th/9711200 [hep-th]].

\bibitem{Witten:1998qj}
E.~Witten,
``Anti de Sitter space and holography,''
Adv.\ Theor.\ Math.\ Phys.\ \textbf{2}, 253-291 (1998)
[arXiv:hep-th/9802150 [hep-th]].

\bibitem{CasalderreySolana:2011us}
J.~Casalderrey-Solana, H.~Liu, D.~Mateos, K.~Rajagopal and U.~Achim~Wiedemann,
``Gauge/String Duality, Hot QCD and Heavy Ion Collisions,''
Cambridge University Press, 2014, ISBN 978-1-009-40350-4, 978-1-009-40349-8, 978-1-009-40352-8, 978-1-139-13674-7
[arXiv:1101.0618 [hep-th]].

\bibitem{Nishioka:2009un}
T.~Nishioka, S.~Ryu and T.~Takayanagi,
``Holographic Entanglement Entropy: An Overview,''
J.\ Phys.\ A \textbf{42}, 504008 (2009)
[arXiv:0905.0932 [hep-th]].

\bibitem{Nishioka:2018khk}
T.~Nishioka,
``Entanglement entropy: holography and renormalization group,''
Rev.\ Mod.\ Phys.\ \textbf{90}, no.3, 035007 (2018)
[arXiv:1801.10352 [hep-th]].

\bibitem{Ryu:2006bv}
S.~Ryu and T.~Takayanagi,
``Holographic derivation of entanglement entropy from AdS/CFT,''
Phys.\ Rev.\ Lett.\ \textbf{96}, 181602 (2006)
[arXiv:hep-th/0603001 [hep-th]].

\bibitem{Hubeny:2007xt}
V.~E.~Hubeny, M.~Rangamani and T.~Takayanagi,
``A Covariant holographic entanglement entropy proposal,''
JHEP \textbf{07}, 062 (2007)
[arXiv:0705.0016 [hep-th]].

\bibitem{Ryu:2006ef}
S.~Ryu and T.~Takayanagi,
``Aspects of Holographic Entanglement Entropy,''
JHEP \textbf{08}, 045 (2006)
[arXiv:hep-th/0605073 [hep-th]].

\bibitem{Balasubramanian:2011ur}
V.~Balasubramanian, A.~Bernamonti, J.~de~Boer, N.~Copland, B.~Craps, E.~Keski-Vakkuri, B.~Muller, A.~Schafer, M.~Shigemori and W.~Staessens,
``Thermalization of Strongly Coupled Field Theories,''
Phys.\ Rev.\ Lett.\ \textbf{106}, 191601 (2011)
[arXiv:1012.4753 [hep-th]].

\bibitem{Klebanov:2007ws}
I.~R.~Klebanov, D.~Kutasov and A.~Murugan,
``Entanglement as a probe of confinement,''
Nucl.\ Phys.\ B \textbf{796}, 274-293 (2008)
[arXiv:0709.2140 [hep-th]].

\bibitem{DeWolfe:2010ca}
O.~DeWolfe, S.~S.~Gubser and C.~Rosen,
``Dynamic critical phenomena at a holographic critical point,''
Phys.\ Rev.\ D \textbf{84}, 126014 (2011)
[arXiv:1108.2029 [hep-th]].

\bibitem{Ebrahim:2023hru}
H.~Ebrahim and M.~Ahmadpour,
``Holographic entanglement entropy and mutual information in deformed field theories at finite temperature,''
Phys.\ Rev.\ D \textbf{107}, no.8, 086010 (2023)
[arXiv:2301.07242 [hep-th]].

\bibitem{Ebrahim:2018zcv}
H.~Ebrahim and M.~Ali-Akbari,
``Dynamically probing strongly-coupled field theories with critical point,''
Phys.\ Lett.\ B \textbf{783}, 43-50 (2018)
[arXiv:1712.08777 [hep-th]].

\bibitem{AliAkbari:2017pcn}
M.~Ali-Akbari and M.~Lezgi,
``Holographic QCD, entanglement entropy, and critical temperature,''
Phys.\ Rev.\ D \textbf{96}, no.8, 086014 (2017)
[arXiv:1706.04335 [hep-th]].

\bibitem{Ebrahim:2019}
H.~Ebrahim, M.~Asadi and M.~Ali-Akbari,
``Evolution of Holographic Complexity Near Critical Point,''
JHEP \textbf{09}, 023 (2019)
[arXiv:1811.12002 [hep-th]].

\bibitem{Casini:2009sr}
H.~Casini and M.~Huerta,
``Entanglement entropy in free quantum field theory,''
J. Phys. A \textbf{42}, 504007 (2009)
[arXiv:0905.2562 [hep-th]].


\bibitem{Takayanagi:2025ula}
T.~Takayanagi,
``Essay: Emergent Holographic Spacetime from Quantum Information,''
Phys. Rev. Lett. \textbf{134}, no.24, 240001 (2025)
[arXiv:2506.06595 [hep-th]].

\bibitem{Doi:2023zyl}
K.~Doi, J.~Harper, A.~Mollabashi, T.~Takayanagi and Y.~Taki,
``Pseudoeentropy in dS/CFT and Timelike Entanglement Entropy,''
Phys.\ Rev.\ Lett.\ \textbf{130}, no.3, 031601 (2023)
[arXiv:2210.09457 [hep-th]].

\bibitem{Doi:2023zyl2}
K.~Doi, J.~Harper, A.~Mollabashi, T.~Takayanagi and Y.~Taki,
``Timelike entanglement entropy,''
JHEP \textbf{05}, 052 (2023)
[arXiv:2302.11695 [hep-th]].

\bibitem{Heller:2024}
M.~P.~Heller, F.~Ori and A.~Serantes,
``Geometric Interpretation of Timelike Entanglement Entropy,''
Phys.\ Rev.\ Lett.\ \textbf{134}, no.13, 131601 (2025)
[arXiv:2408.15752 [hep-th]].

\bibitem{Jena:2025}
S.~S.~Jena and S.~Mahapatra,
``A note on the holographic time-like entanglement entropy in Lifshitz theory,''
JHEP \textbf{01}, 055 (2025)
[arXiv:2410.00384 [hep-th]].

\bibitem{Nakata:2020rsh}
Y.~Nakata, T.~Takayanagi, Y.~Taki, K.~Tamaoka and Z.~Wei,
``New holographic generalization of entanglement entropy,''
Phys.\ Rev.\ D \textbf{103}, no.2, 026005 (2021)
[arXiv:2005.13801 [hep-th]].

\bibitem{Caputa:2024zxf}
P.~Caputa, B.~Chen, T.~Takayanagi and T.~Tsuda,
``Thermal pseudo-entropy,''
JHEP \textbf{01}, 003 (2025)
[arXiv:2411.08948 [hep-th]].

\bibitem{Afrasiar:2025b}
M.~Afrasiar, J.~K.~Basak and K.~Y.~Kim,
``Aspects of holographic timelike entanglement entropy in black hole backgrounds,''
[arXiv:2512.21327 [hep-th]].

\bibitem{AliAkbari:2026}
M.~Ali-Akbari,
``Probe Dependence of the Imaginary Part of HTEE,''
[arXiv:2608.04599 [hep-th]].

\bibitem{Goki:2026}
M.~M.~D.~Goki and M.~Ali-Akbari,
``On holographic time-like entanglement entropy,''
Eur.\ Phys.\ J.\ C \textbf{86}, no.6, 664 (2026)
[arXiv:2601.17810 [hep-th]].

\bibitem{Sahay:2010rq}
A.~Sahay, T.~Sarkar and G.~Sengupta,
``On The Phase Structure and Thermodynamic Geometry of R-Charged Black Holes,''
JHEP \textbf{11}, 125 (2010)
[arXiv:1009.2236 [hep-th]].


\bibitem{DeWolfe:2010ca2}
O.~DeWolfe, S.~S.~Gubser and C.~Rosen,
``A holographic critical point,''
Phys.\ Rev.\ D \textbf{83}, 086005 (2011)
[arXiv:1012.1864 [hep-th]].

\bibitem{Amrahi:2020jqz}
B.~Amrahi, M.~Ali-Akbari and M.~Asadi,
``Holographic entanglement of purification near a critical point,''
Eur.\ Phys.\ J.\ C \textbf{80}, no.12, 1152 (2020)
[arXiv:2004.02856 [hep-th]].

\end{thebibliography}
\end{document}